\documentclass[aps,prd,reprint,superscriptaddress]{revtex4-1}

\usepackage{hyperref}
\usepackage{graphicx}
\usepackage{enumerate}
\hypersetup{
    pdfnewwindow=true,      
    colorlinks=true,       
    linkcolor=blue,          
    citecolor=blue,        
    filecolor=blue,      
    urlcolor=blue           
}

\usepackage{subfigure}
\usepackage{amsmath}
\usepackage{amssymb}
\usepackage{amsfonts}
\usepackage{float}

\usepackage{color}

\providecommand{\aap}[0]{A\&A }

\providecommand{\astropartphys}[0]{Astropart. Phys. } 

\providecommand{\apj}[0]{ApJ }
\providecommand{\apjl}[0]{ApJ Lett. }
\providecommand{\apjs}[0]{ApJ Supp. Ser. }

\providecommand{\cqg}[0]{CQG}

\providecommand{\jcap}[0]{J. Cosmol. Astropart. P. }
\providecommand{\mnras}[0]{MNRAS }

\providecommand{\nat}[0]{Nature }
\providecommand{\physrep}[0]{Phys. Rep. }
\providecommand{\prd}{PRD }
\providecommand{\prl}[0]{PRL }
\providecommand{\rmp}[0]{Rev. Mod. Phys. } %
\providecommand{\repprogphys}[0]{Rep. Prog. Phys. }

\def\beq{\begin{equation}}
\def\eeq{\end{equation}}

\begin{document}

\title{Prospects of Establishing the Origin of Cosmic Neutrinos using Source Catalogs}

\author{I. Bartos}
\email{ibartos@phys.columbia.edu}
\affiliation{Department of Physics, Columbia University, New York, NY 10027, USA}
\author{M. Ahrens}
\affiliation{Oskar Klein Centre \& Dept. of Physics, Stockholm University, SE-10691 Stockholm, Sweden}
\author{C. Finley}
\affiliation{Oskar Klein Centre \& Dept. of Physics, Stockholm University, SE-10691 Stockholm, Sweden}
\author{S. M\'arka}
\affiliation{Department of Physics, Columbia University, New York, NY 10027, USA}

\begin{abstract}
The cosmic neutrino flux recently discovered by IceCube will be instrumental in probing the highest-energy astrophysical processes. Nevertheless, the origin of these neutrinos is still unknown. While it would be more straightforward to identify a transient, or galactic source class, finding a population of distant, continuous sources is challenging. We introduce a source-type classification technique that incorporates all available information from catalogs of source candidates. We show that IceCube-Gen2 can statistically establish the origin of cosmic neutrinos even for the most challenging source populations--starburst galaxies, AGN or galaxy clusters, if neutrino track directions can be reconstructed with a precision $\sim0.3^\circ$. We further show that source catalog out to $\sim100$\,Mpc can be sufficient for the most challenging source types, allowing for more straightforward source surveys. We also characterize the role of detector properties, namely angular resolution, size and veto power in order to understand the effects of IceCube-Gen2's design specifics.
\end{abstract}


\maketitle

\section{Introduction}

Cosmic high-energy neutrinos provide a unique probe of energetic astrophysical processes. They carry information on the origin of cosmic rays \cite{2002RPPh...65.1025H}, the nature of particle acceleration in cosmic explosions \cite{2014arXiv1412.5106I}, and the properties of these explosions \cite{2003PhRvD..68h3001R,2012PhRvD..86h3007B}. They can escape dense environments and travel through cosmic distances, allowing for a direct view of emission sites hidden from high-energy electromagnetic observations.


IceCube has recently detected high-energy neutrinos of astrophysical origin, with a detection rate of $\sim 10$\,yr$^{-1}$ for neutrino energies $\gtrsim 10^2$TeV \cite{2015PhRvD..91b2001A,2015PhRvL.115h1102A,2015ApJ...809...98A}. Nevertheless, the cosmic source population producing this neutrino flux is currently unknown. A planned upgrade of IceCube, called IceCube-Gen2 \cite{2014arXiv1412.5106I}, is expected to improve neutrino detection rate by a factor of $\sim10$, enhancing the prospects of identifying the origin and emission mechanism of high-energy neutrinos. Another proposed neutrino detector in the Mediterranean, KM3NeT \cite{KM3NeT}, will be comparably sensitive, and will be sensitive in directions complementary to IceCube.


There are multiple search strategies being considered that aim to determine the origin of the cosmic neutrinos.
\begin{enumerate}[(i)]
\item \emph{Neutrino-only}: spatial or temporal clustering of neutrinos alone could be used to identify energetic nearby sources, even without any further information on potential source candidates \cite{2014ApJ...796..109A,2015ApJ...807...46A,2015APh....66...39A,2016arXiv160908027F,PhysRevD.94.103006}. No such clustering has been detected so far, which was used to constrain rare, bright source populations \cite{2014PhRvD..90d3005A}. These analyses are essentially model-independent, and they do not require multimessenger observations, but they are less sensitive than those that utilize additional multimessenger information.
\item \emph{Spectral information} from the detected neutrinos in comparison with other observed astrophysical fluxes, such as cosmic rays \cite{1997PhRvL..78.2292W} or GeV gamma rays \cite{2013PhRvD..88l1301M}, can help determine the source of origin. Neutrino spectral features may also be indicative of the source's environment \cite{2006JCAP...05..003L,2015arXiv150900983B}. It is, nevertheless, difficult to arrive at definite conclusions solely by such comparisons.
\item \emph{Multimessenger searches}: neutrinos can be associated with a source population observed through other messengers \cite{2008PhR...458..173B,2011PhRvL.107y1101B,2012Natur.484..351A,2013RvMP...85.1401A,2014ApJ...796..109A,PhysRevD.94.103006}. This strategy has the advantage of integrating additional information in the search. Ultimately, identifying the neutrino's origin will likely require such an association.
\end{enumerate}

So far, most of the searches using astrophysical source catalogs, e.g., using gamma-ray bursts (GRBs; \cite{2015ApJ...805L...5A}) and gravitational wave candidates \cite{2014PhRvD..90j2002A,2016PhRvD..93l2010A}, focused on relatively rare sources with limited neutrino background \cite{2014PhRvD..90j1301B}, or a limited subset of more common source types \cite{2014ApJ...796..109A}. No correlation has been identified in these searches. The neutrino directional distribution also seems to indicate that the bulk of the detected neutrinos originates outside of the Milky Way. Recently, Ahlers \& Halzen \cite{2014PhRvD..90d3005A} considered the general case of using the locations of the brightest sources to enhance search sensitivity to neutrino multiplets, making an important first step and characterizing the expected sensitivity. Their results indicate that, in contrast to the case of transient sources, the identification of multiplets from a population of continuous sources will be challenging, even with an extended observation period with IceCube-Gen2, except for rare source types.

Search strategies are relatively straightforward for transient sources for which the limited duration efficiently reduces the background. Similarly, for rare, bright sources, looking at the nearest, brightest sources can already be informative \cite{2014PhRvD..90d3005A}. However, for weaker, continuous sources, the utilization of source catalogs is not straightforward since the majority of the detected neutrinos were emitted at distances at which the source population is homogeneous on the scale of the neutrino directional uncertainty. This means that distant sources do not help with the association. On the other hand, it is advantageous to use more than just the closest sources in correlation with the detected neutrinos, since (i) the closest (non-galactic) sources likely emit only a negligible fraction of the total neutrino flux, and (ii) source distributions are sufficiently anisotropic out to distances containing a large number of sources, even for continuous source models with relatively high source densities.

This paper explores the prospects of using catalogs of astrophysical sources to search for the origin of high-energy neutrinos. Our goal is to devise and evaluate a search strategy that efficiently utilizes source catalogs. We derive the prospects of using such a search strategy for a number of possible source models, namely (i) starburst galaxies, (ii) the cores of active galactic nuclei (AGN), (iii) blazars and (iv) galaxy clusters (GCs). In particular we determine the angular precision of neutrino detectors that would allow for the identification of different source types.

The paper is organized as follows. We present the search strategy in Section \ref{section:searchstrategy}. We then discuss our results on search sensitivity in Section \ref{section:results}. In particular we use Monte Carlo simulations to explore the role of the detector angular precision and the catalog distance range. We summarize the results and present our conclusions in Section \ref{section:conclusion}.

\section{Search strategy}
\label{section:searchstrategy}

\subsection{Search distance threshold}

Our goal is to take advantage of the anisotropic distribution of astrophysical sources, and correlate the sources' directional variation with the directions of the detected neutrinos. It is clear that at large enough distances, the source population is quasi-homogeneous, rendering the position of these distant sources uninformative. We can therefore constrain the search to sources within a threshold distance $d_{\rm th}$ (or threshold redshift $z_{\rm th}$).

Constraining the search to neutrinos emitted from known sources within $d_{\rm th}$ means, in effect, that neutrinos originating in these known sources will be the \emph{signal} neutrinos, while \emph{background} neutrinos will be those (i) originating in the atmosphere, (ii) from sources beyond $d_{\rm th}$, and (iii) from un-cataloged sources within $d_{\rm th}$.

\subsection{Only track events will be used}

The uncertainty of the reconstructed direction of neutrinos varies widely. In IceCube, for instance, events due to charged-current muon neutrino interactions produce tracks with directional uncertainty $\lesssim 1^\circ$, while charged-current electron neutrino interactions (as well as all neutral current interactions) produce shower event topologies with directional uncertainty $\sim 10^\circ$ \cite{2013Sci...342E...1I}.

With these rates and angular uncertainties, it is beneficial to use only track events in the search for directional correlation with a source population.

\subsection{Method}
\label{section:method}

Here we introduce our search technique, for continuous neutrino sources. It is straightforward to generalize the strategy to transients see below).

Let the total number of detected neutrino candidates be $N_{\rm total}$ within an observation period $T_{\rm obs}$. Assume that we know the expected number $N_{\rm atm}$ of atmospheric neutrinos within the total from atmospheric contribution models. For our threshold distance $d_{\rm th}$, we can calculate the expected fraction $f_{\rm d}$ of astrophysical neutrinos that originate from within $d_{\rm th}$ (see Section \ref{section:fraction}).
%
%
%
Our expected signal ($N_{\rm s}$) and background ($N_{\rm b}$) neutrinos will then be be
\begin{eqnarray}
N_{\rm s} &=& f_{\rm d}\,(N_{\rm total} - N_{\rm atm}), \\
N_{\rm b} &=& N_{\rm total} - N_{\rm s}.
\label{eq:N}
\end{eqnarray}

For a detected neutrino $i$ with reconstructed sky location $\Omega_{\rm i}$ and location uncertainty $\psi_{\rm i}$, the neutrino's volume of possible origin $V_{\rm i}$ can be approximated with a cone with height $d_{\rm th}$ and base radius $\psi_{\rm i} d_{\rm th}$, oriented along $\Omega_{\rm i}$. We calculate the expected total neutrino flux from sources within $V_{\rm i}$:
\begin{equation}
\mathcal{F}_{\rm \nu,i} = \frac{1}{4\pi}\sum_j L_{\rm \nu,j}\,d_{\rm j}^{-2},
\label{eq:flux}
\end{equation}
where $L_{\rm \nu,j}$ is the neutrino luminosity of source $j$, and the sum is over the sources within $V_{\rm i}$. See Fig. \ref{figure:TSdist} for a sample distribution of $\mathcal{F}_{\rm \nu,i}$ for signal and background neutrinos. Neutrino luminosity values depend on the source model considered. We adopt identical luminosities for each source within a given source class. This is a simplifying assumption that, nevertheless, makes our results more conservative. Selecting instead a distribution of non-equal luminosities would, on average, increase the flux of the brightest sources, for which both distance and luminosity are favorable. Since we are considering sources within $\lesssim200$\,Mpc, redshift evolution will not affect this picture.

The assumption of uniform source luminosity is also practical, given that individual source luminosities are unknown, and are unlikely to become available in the next years unless some individual neutrino emitters are clearly detected.

To demonstrate why our assumption of uniform source luminosities is conservative, imagine the simple case in which we have a set of cataloged source candidates, all with equal weight. If we now take one source candidate $A$ and place it onto candidate $B$, we effectively double $B$'s weight and make $A$'s weight zero. It is clear that this change will increase anisotropy. The same argument can be made with fractional changes of weights, or similarly, with a non-uniform weight distribution.

Beyond this assumed uniformity, the actual numerical values of $L_{\rm \nu,j}$ are not important as they will not affect our results, since only their relative weight matters in determining spacial correlation.

The background probability density $\mathcal{B}_{\rm i}$ of neutrino $i$ will be taken to be uniform over the sensitive sky region of the detector:
\begin{equation}
\mathcal{B}_{\rm i} = \mbox{const.}
\end{equation}
Here, for simplicity, we do not make use of the energy of the neutrino. The reconstructed neutrino energy would be straightforward to incorporate into the signal and background likelihoods (see, e.g.,\cite{2008APh....29..299B}), which would further improve the search sensitivity.

We define the signal probability density $\mathcal{S}_{\rm i}$ to be proportional to the expected neutrino flux from sources within $V_{\rm i}$:
\begin{equation}
\mathcal{S}_{\rm i} \propto \frac{\mathcal{F}_{\rm \nu,i}}{\sin^2(\psi_{\rm i})}.
\end{equation}

For the ensemble of observed neutrinos, we combine the signal and background densities, to obtain the likelihood
\begin{equation}
\mathcal{L}(N_{\rm s}, N_{\rm total}) = \prod_{i}\left(\frac{N_{\rm s}}{N_{\rm total}}\mathcal{S}_{\rm i} + \frac{N_{\rm b}}{N_{\rm total}}\mathcal{B}_{\rm i}\right),
\label{eq:likelihood}
\end{equation}
where the product is over all detected neutrinos during the observation period. While $N_{\rm s}$ is not known precisely, it can be estimated using $N_{\rm total}$, $N_{\rm atm}$ and $f_{\rm d}$ (see Eqs. \ref{eq:N}). In general, one could maximize $\mathcal{L}(N_{\rm s}, N_{\rm total})$ as a function of $N_{\rm s}$ as use this value for the calculation (see, e.g.,\cite{2008APh....29..299B}). For simplicity, we will omit this maximization.

The test statistic of the observed set of neutrinos will be the likelihood ratio
\begin{equation}
\lambda = 2 \log\left[\frac{\mathcal{L}(N_{\rm s}, N_{\rm total})}{\mathcal{L}(0, N_{\rm total})}\right].
\label{eq:likelihoodratio}
\end{equation}

The significance of the ensemble of observed neutrinos is determined by comparing the observed $\lambda$ value to the background distribution $P_{\rm bg}(\lambda)$, determined using Monte Carlo simulations. The p-value of the observed ensemble will be defined as
\begin{equation}
p_{\lambda} = \int_{\lambda}^{\infty} P_{\rm bg}(\lambda')d\lambda'.
\end{equation}

Here we also mention how the above method, presented for the case of continuous sources, compares to the transient case. For the purposes of the search described below, a transient rate density $R_{\rm transient}$ is equivalent to an effective continuous source number density $n_{\rm continuous,effective}\approx R_{\rm transient}T_{\rm transient}$, where $T_{\rm transient}$ is the characteristic duration of the transient, and assuming $R_{\rm transient}T_{\rm transient}\gg1$. Other aspects of the method are unchanged.

\subsection{Astrophysical source types}
\label{section:sourcetypes}

In the following, we will consider four astrophysical source types, namely starburst galaxies, AGN, blazars and GCs. These are the primary considered continuous source candidates for the detected cosmic neutrino flux. We will further assume that all cosmic neutrinos originate from the same source type.

We follow \cite{2014PhRvD..90d3005A} and adopt local number densities $\rho_{\rm starburst}=10^{-4}$\,Mpc$^{-3}$, the range $\rho_{\rm agn}=10^{-5}-10^{-4}$\,Mpc$^{-3}$, and $\rho_{\rm blazar}=10^{-9}$\,Mpc$^{-3}$ for starburst galaxies, AGN and blazars, respectively. For GCs, we follow \cite{2016PhRvD..94j3006M} and adopt $\rho_{\rm gc}=5\times10^{-5}$\,Mpc$^{-3}$.

For simplicity, we assume that $L_{\rm \nu}$ is identical for all sources within a given source type. This is a conservative assumption; taking into account the variation in luminosity would increase anisotropy, and therefore make the search more sensitive. We expect that typical source catalogs will have limited information on source strength. Nevertheless, if such information is available, it is straightforward to incorporate it in the analysis by simply using the derived luminosity values, which in turn will further improve our sensitivity.

\subsection{Fraction of neutrinos from within threshold distance}
\label{section:fraction}

To find the number of detected astrophysical neutrinos, we need to estimate the fraction of the neutrino flux at Earth that originates from the cataloged sources. Only these neutrinos will correlate with the anisotropy of the catalog. We first determine the fraction $f_{\rm d}(z_{\rm th})$ of astrophysical neutrinos that originate from within a threshold redshift $z_{\rm th}$, where $z_{\rm th}$ corresponds to threshold distance $d_{\rm th}$.

The total neutrino particle flux $\Phi_\nu \equiv dN_\nu / d\varepsilon_\nu'$ per unit energy band from sources within redshift $z_{\rm max}$ can be written as \cite{2014PhRvD..89h3004L}
\begin{equation}
\Phi_\nu (\varepsilon_\nu',z_{\rm max}) = \frac{1}{4\pi}\int_{0}^{z_{\rm max}}\rho(z)\phi_\nu[(1+z)\varepsilon_\nu']\frac{cdz}{H(z)}.
\end{equation}
Here, $\varepsilon_\nu'$ is the neutrino energy at the observer, $\phi_\nu$ is the average neutrino spectrum from a source, and $H(z)=H_0\sqrt(\Omega_{\rm M}(1+z)^3+\Omega_\Lambda)$ is the Hubble parameter. We adopt $H_0=67.3$\,km\,s$^{-1}$Mpc$^{-1}$, $\Omega_{\rm M}=0.315$ and $\Omega_{\Lambda}=0.685$ \cite{2014A&A...571A..16P}. We assume a power-law source spectrum $\phi_\nu\propto\varepsilon_\nu^\gamma$, for which $\Phi_\nu(\varepsilon_\nu',z_{\rm max}) = \Phi_\nu(z_{\rm max})$ will be independent of the neutrino energy as long as the energy is far enough from a spectral cut-off energy. The fraction of neutrinos coming from sources within the threshold $z_{\rm th}$ is therefore $f_{\rm d}(z_{\rm th}) = \Phi_\nu(z_{\rm th})/\Phi_\nu(\infty)$.

We calculate $f_{\rm d}(z_{\rm th})$ for different source models relevant to plausible neutrino emission scenarios. For the case of starburst galaxies, we assume that neutrino emission tracks the cosmic star formation rate (SFR). We adopt the SFR evolution from \cite{2008MNRAS.388.1487L}, taking power indices $\alpha_1=3.3$, $\alpha_2=0.055$ and $\alpha_3=-4.46$, and breaks $z_1=0.993$ and $z_2=3.8$. For AGN and blazars, we assume that neutrino emission tracks the hard X-ray emission of AGN (e.g., \cite{2014PhRvD..90d3005A}), and adopt the evolution of the hard X-ray luminosity function of AGN from \cite{2010MNRAS.401.2531A}. For GCs, we adopt an evolution of $(1+z)^{3}$ \cite{2016PhRvD..94j3006M}. For all cases, to good approximation we obtain the linear relationship $f_{\rm d}(d_{\rm th}) = f_{0} (d_{\rm th} / 1\,\mbox{Mpc})$ for $d_{\rm th}\lesssim 1$\,Gpc. This is not surprising, as we expect such linear relationship for the no-evolution case as well. The multiplicative factors we find are therefore $f_{0}^{\rm starburst} = f_{0}^{\rm sfr} \approx9\times10^{-5}$, $f_{0}^{\rm agn} \approx 6 \times10^{-5}$, $f_{0}^{\rm blazar} \approx f_{0}^{\rm agn}$, and $f_{0}^{\rm gc} \approx 5 \times10^{-5}$.

\begin{figure}
\resizebox{0.5\textwidth}{!}{\includegraphics{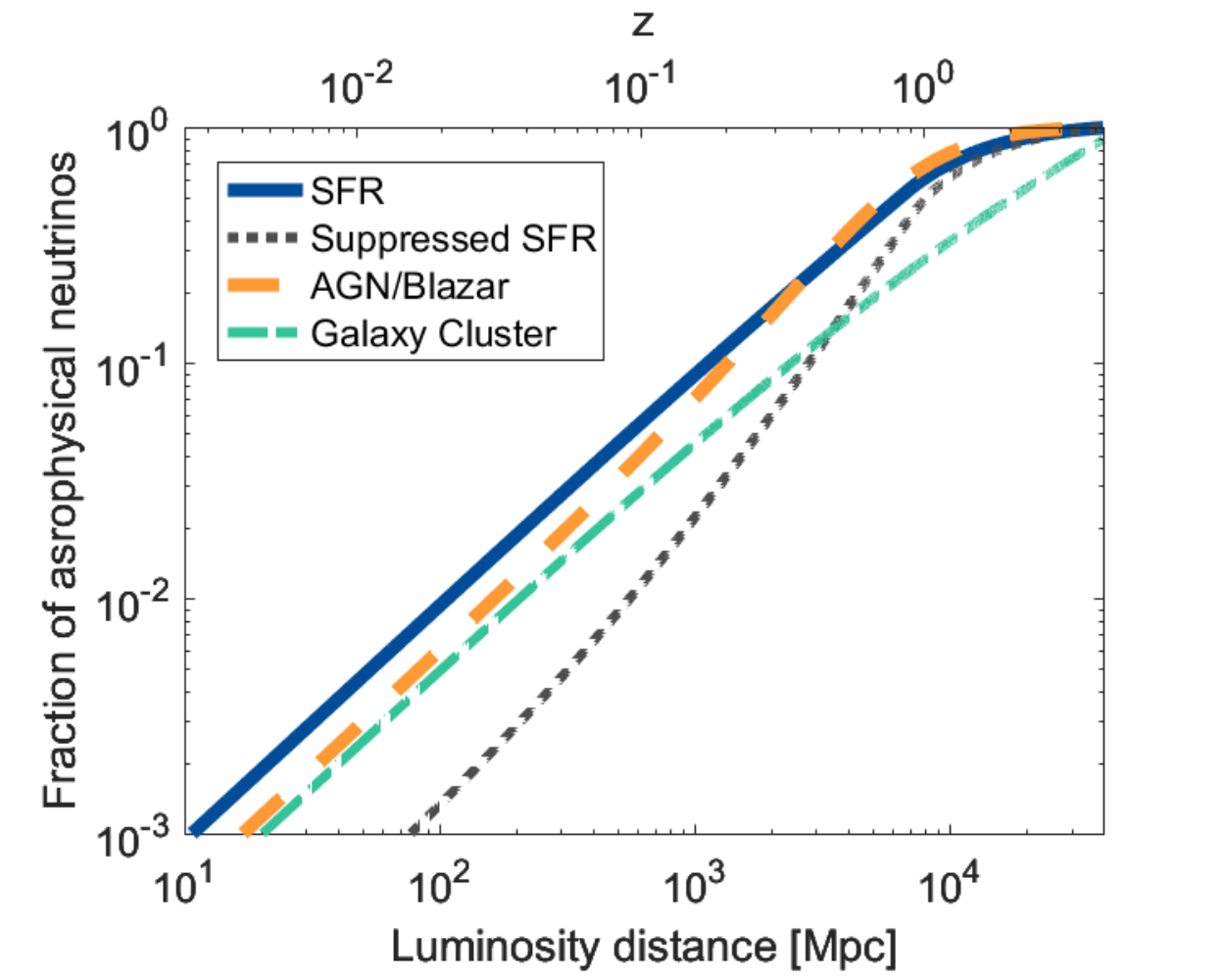}}
\caption{Fraction of astrophysical neutrinos detected from sources within a luminosity distance, as a function of the luminosity distance, for different cosmic evolution models (see legend and Section \ref{section:fraction}). For comparison the top axis shows the corresponding redshift.}
\label{figure:NeutrinoGalaxy}
\end{figure}

Starburst galaxies' neutrino production may evolve differently with redshift than the SFR, as indicated by the evolution of their observed infrared emission \cite{2006astro.ph..8699T,2014PhRvD..90d3005A}. Using local far-infrared and radio luminosity densities observed from starburst galaxies, Thompson et al. \cite{2006astro.ph..8699T} estimates their contribution to star formation to be $\sim10\%$, which can grow to $100\%$ at high redshifts. Introducing a corresponding additional factor of $\mbox{min}(0.1+0.9z,1)$ to the evolution of star formation rate for starburst galaxies, we find that this locally suppressed contribution to neutrino production will give $f_{0}^{\rm suppressed} \approx 1.3\times10^{-5}$. Nevertheless, more recent measurements of the infrared luminosity function by Herschel \cite{2013MNRAS.432...23G} found that the evolution of the luminosity of starburst galaxies is not markedly different from the evolution of SFR \cite{2014JCAP...09..043T}. We will therefore adopt $f_{0}^{\rm starburst}$ for the analysis below. In Section \ref{sec:suppression} we separately comment on how these results would change if starburst contribution is indeed locally suppressed.

\subsection{Comparison to previous work}

We briefly outline the methodology of some previous searches for the origin of cosmic neutrinos to present our work in comparison.

\section{Results}
\label{section:results}

\subsection{Neutrino detection rate}
\label{section:detectionrate}

For our results we need to estimate the number of detected astrophysical and background neutrino track events for IceCube-Gen2.  We will assume an astrophysical flux of $0.9\times 10^{-18} (E/100\,\mbox{TeV})^{-2.13}\,$GeV$^{-1}$sr$^{-1}$cm$^{-2}$s$^{-1}$ as measured in \cite{2016arXiv160708006I}.   In the future IceCube-Gen2 built in its Sunflower configuration with 240 strings \cite{2014arXiv1412.5106I}, the estimates above a threshold of $\sim30$\,TeV for the rate of astrophysical neutrino track events $R_{\rm \nu,ast}$ and background atmospheric events $R_{\rm \nu,atm}$ are 36\,yr$^{-1}$ for each, roughly 10 times the current IceCube rate \cite{2013Sci...342E...1I}.


We will alternatively consider a broader, low-threshold search ($\gtrsim1$\,TeV) with IceCube-Gen2 in its Sunflower configuration with 240 strings. This search is expected to yield $R_{\rm \nu,ast}\approx200$\,yr$^{-1}$ and $R_{\rm \nu,atm}\approx15000$\,yr$^{-1}$.


\subsection{Monte Carlo simulation}
\label{section:MC}

We carry out Monte Carlo simulations to determine expected neutrino flux anisotropy for different source populations, and from this to determine the probability of identifying a source type as the origin of the astrophysical neutrinos with sufficiently high significance. This probability will depend on the expected total number ($N_{\rm total} - N_{\rm atm}$) of detected astrophysical neutrinos, the expected number ($N_{\rm atm}$) of detected atmospheric neutrinos, the source number density $\rho$, the source catalog threshold distance $d_{\rm th}$ number, neutrino directional uncertainty $\psi$, and significance (p-value) threshold $p_0$.

For the analysis below, we consider all neutrinos having the same angular uncertainty $\psi$. For simplicity we will assume that the true neutrino direction is uniformly distributed within $\psi$ of the reconstructed direction. We will be interested in how the search sensitivity depends on $\psi$.

For a given set of the parameters listed above, we carry out Monte Carlo realizations as follows. We first generate a set of $>10^7$ neutrinos. For each neutrino, we randomly generate a number of sources that are directionally coincident with the neutrino. The number of sources is drawn from a Poisson distribution with $V\rho$ mean, with $V=\pi d_{\rm th}^3 \sin^2(\psi)/3$. For each source, we randomly select a distance using uniform distribution within $V$. We calculate the expected neutrino flux $\mathcal{F}$ using Eq. \ref{eq:flux}, using constant source luminosity.

\begin{figure}
\resizebox{0.475\textwidth}{!}{\includegraphics{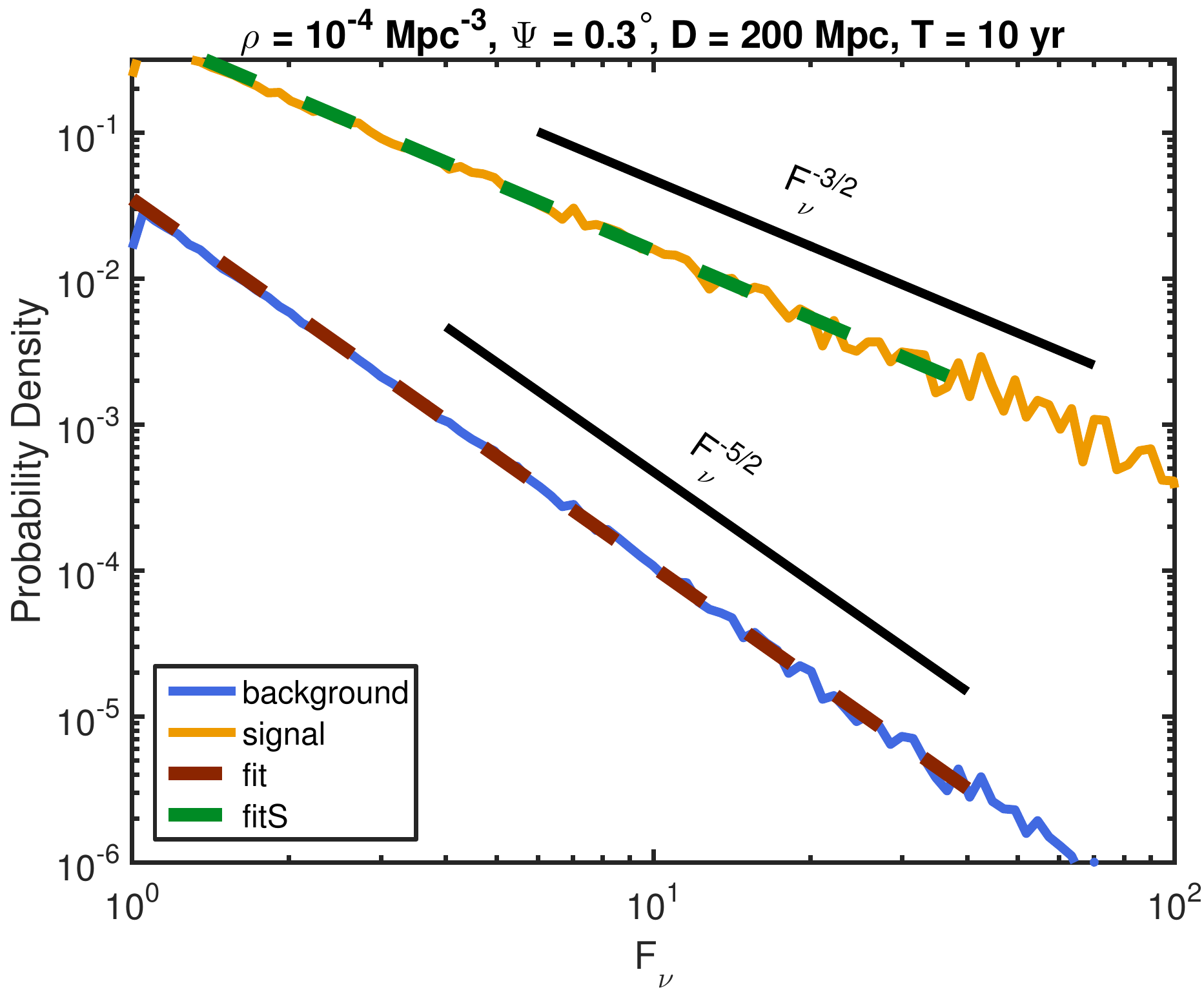}}
\caption{Simulated distribution of neutrino flux $\mathcal{F}_{\rm \nu}$ corresponding to signal and background neutrinos. This example shows the densities for $\psi = 0.3^\circ$, $d_{\rm th}=200$\,Mpc, for starburst galaxies. The dashed red line shows a power-law fit on the background density's  tail. The black solid lines show power-law slopes with $\mathcal{F_{\nu}}^{-3/2}$ and $\mathcal{F_{\nu}}^{-5/2}$, the theoretical expectations for the signal and neutrino models, respectively, for the case in which the expected number of sources within $d_{\rm th}$ coincident with a neutrino is $\ll1$.}
\label{figure:TSdist}
\end{figure}

Next, we calculate the probability density of $\mathcal{F}_{\nu}$ corresponding to signal and background neutrinos from the generated set of neutrinos. The probability of signal neutrinos having a flux between $\mathcal{F}_{\nu}$ and $\mathcal{F}_{\nu}+d\mathcal{F}_{\nu}$ will be proportional to the number of randomly generated neutrinos falling within this flux range, times $\mathcal{F}_{\nu}$. The same probability for background neutrinos will be simply proportional to the number of randomly generated neutrinos falling within this flux range. An example of the signal and background $\mathcal{F}_{\nu}$ probability densities are shown in Fig. \ref{figure:TSdist}. We see that the tail distributions agree with the theoretically expected $\mathcal{F}_{\nu}^{-3/2}$ and $\mathcal{F}_{\nu}^{-5/2}$ for the signal and background cases, respectively.

To better estimate the probability density of $\mathcal{F}_{\nu}$ corresponding to larger $\mathcal{F}$ values for which the sampling is sparse, we fit a power law on the tail of the background $\mathcal{F}_{\nu}$ probability density. An example of this fit is shown in Fig. \ref{figure:TSdist}.

For a realization of the Monte Carlo simulation, we generate signal and background neutrino numbers $N_{\rm s}$ and $N_{\rm b}$, respectively. Both numbers are drawn from a Poisson distribution with their expected means (see Eq. \ref{eq:N} and the parameters above). For these signal and background neutrinos, we randomly select corresponding $\mathcal{F}_{\nu}$ values drawn from the probability densities of $\mathcal{F}$ as described above. We then calculate the test statistic $\lambda$ for the realization, using Eqs. \ref{eq:likelihoodratio} and \ref{eq:likelihood}. We generate $10^5$ such realizations to obtain $P_{\rm s}(\lambda)$, the distribution of $\lambda$, assuming the presence of an astrophysical signal originating from the source model being tested. Note that, for Eq. \ref{eq:likelihood}, we use the expected $N_{\rm s}$ and $N_{\rm b}$, i.e. not the actual ones drawn from the Poisson distribution, as those are assumed to be unknown.

\begin{figure}
\resizebox{0.475\textwidth}{!}{\includegraphics{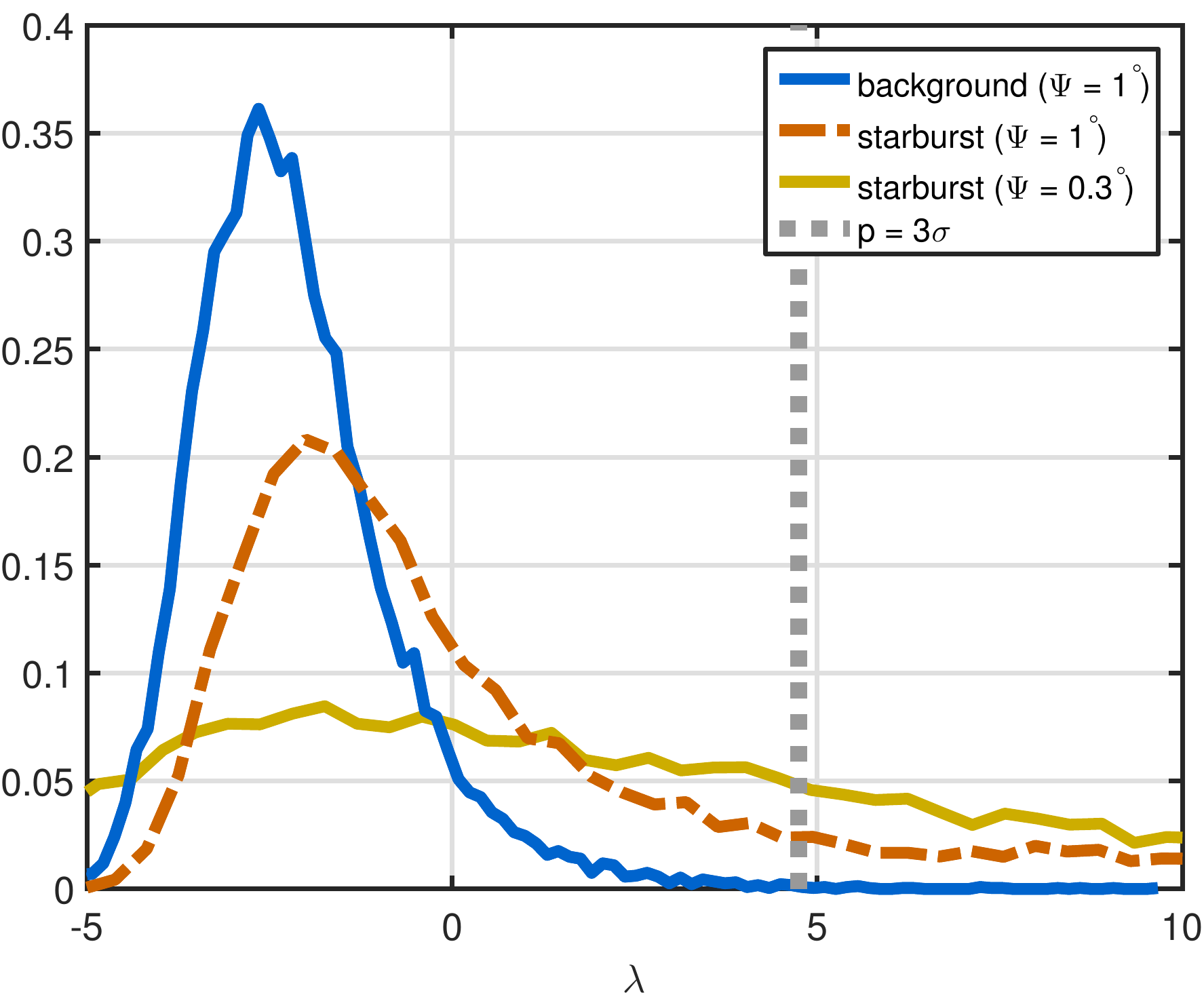}}
\caption{Simulated distributions of likelihood ratio $\lambda$ for the signal+background and background-only models. This example shows the distributions for $d_{\rm th}=200$\,Mpc, $T_{\rm obs}=10$\,yr, for starburst galaxies, and with angular uncertainties specified in the legend. The vertical dashed line shows the likelihood ratio corresponding to $p=3\sigma$ p-value from the background distribution.}
\label{figure:likelihoodratio}
\end{figure}

We also generate $10^5$ realizations for a background only model, i.e. for which the expected number of background neutrinos is $N_{\rm b}' = N_{\rm s} + N_{\rm b}$, and the expected number of signal neutrinos is $N_{\rm s}' = 0$. This will yield $P_{\rm b}(\lambda)$, that is the distribution of $\lambda$ assuming no astrophysical signal is present correlated with the source model being tested. Examples of the resulting likelihood ratio probability densities for the signal+background and background-only models are shown in Fig. \ref{figure:likelihoodratio}.

We characterize a set of parameters by the probability that a random realization of signal+background neutrinos produces $\lambda$ that is greater than $1-p_0$ fraction of the background-only $\lambda$ values. In other words, we characterize the sensitivity by the probability $\mathcal{P}(p_0)$ of obtaining a result from a signal+background model with p-value $\leq p_0$.

\subsection{Optimal threshold distance for source catalog}

\begin{figure}
\resizebox{0.5\textwidth}{!}{\includegraphics{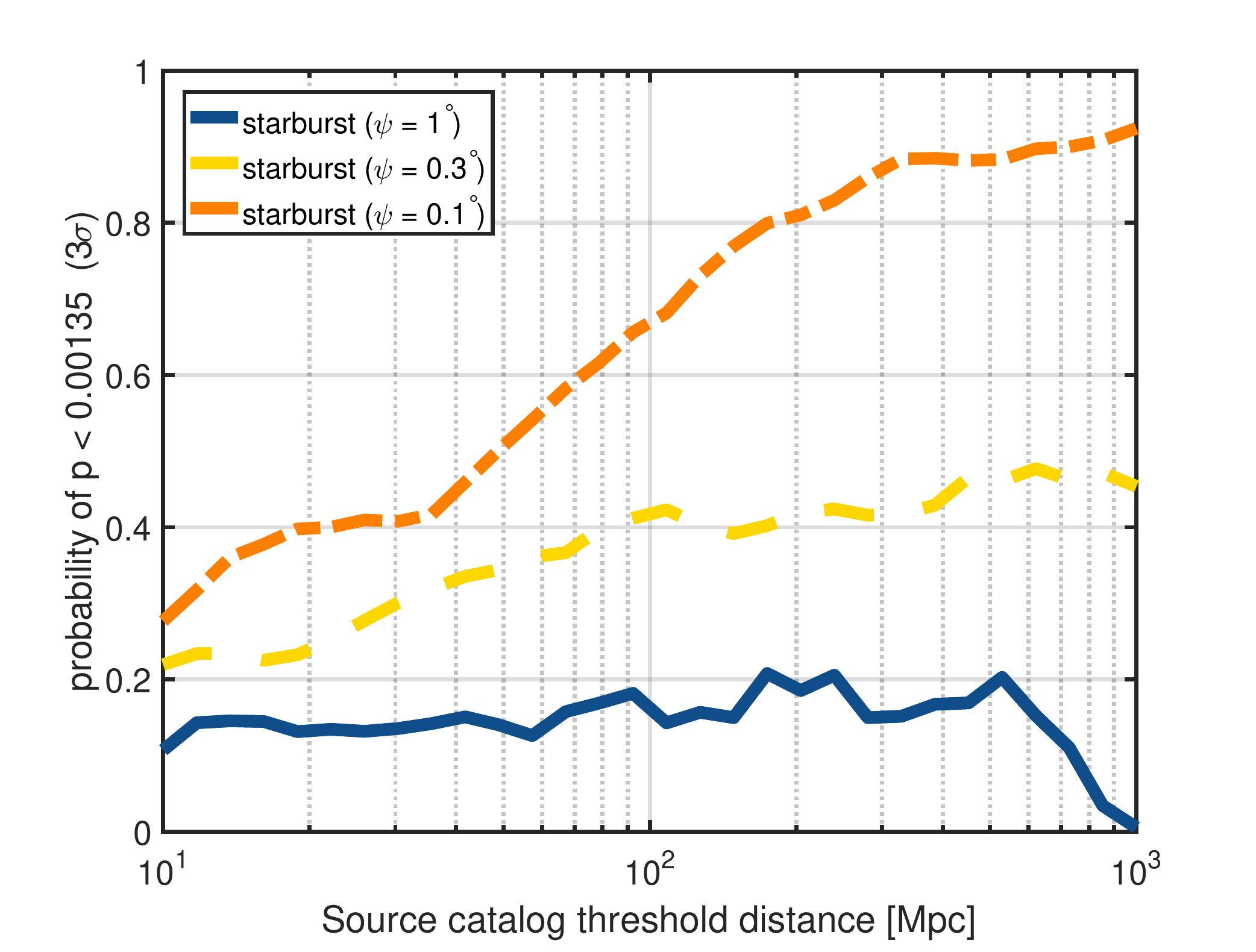}}
\caption{Probability of a signal+background observation having a p-value $\leq p_0=3\sigma=0.00135$ as a function of threshold distance $d_{\rm th}$, for different angular uncertainties (see legend). The results are obtained using the starburst-galaxy model, assuming $T_{\rm obs}=10$\,yr with IceCube-Gen2 for the highest energy ($\geq30$\,TeV) neutrinos.}
\label{figure:distancethreshold}
\end{figure}

The optimal threshold distance for a search depends on the level of anisotropy determined by the source number density, as well as the completeness of the catalog as a function of distance. To characterize the optimal threshold distance for source catalogs, we calculate the sensitivity of the catalog-based search as a function of catalog threshold distance. We use Monte Carlo simulations as described above (Section \ref{section:MC}) to obtain $\mathcal{P}(p_0)$ as a function of $d_{\rm th}$ for different $\psi$ values, using $T_{\rm obs}=10$\,yr with IceCube-Gen2 (Sunflower 240 configuration, 24\,TeV energy threshold). The results are shown in Fig. \ref{figure:distancethreshold} for the case of starburst galaxies. We see that the sensitivity of the catalog-based search increases with increasing $d_{\rm th}$ up to a point at which it saturates. The point of saturation, as well as the maximum sensitivity, scale with $\psi$. This saturation is beyond the 1\,Gpc shown here for the most accurate localization with $0.1^\circ$ uncertainty considered here. This saturation is expected; at larger distances, the average angular source density is $\gg1$, making directional density fluctuations small. This means that sources at large distances will not significantly contribute to sensitivity.

An important consequence of this result is that even a catalog out to $\mathcal{O}(100$\,Mpc$)$ can be close to optimal unless angular uncertainty is small ($\sim0.1^{\circ}$). This is important since source catalog completeness can be significantly reduced for larger distances. Nevertheless, for high directional accuracy, catalogs out to $\gtrsim 200$\,Mpc may help. While catalogs are currently far from complete at 200\,Mpc, we point out that, for a relatively small number of neutrinos, one can construct such a catalog specifically for the direction of the neutrinos (see \cite{2041-8205-801-1-L1} for a similar analysis). This significantly reduces the required observation time, making such a neutrino-specific catalog possible.

It is worth considering here the current and future availability of source catalogs with the above threshold distance in mind. Starburst galaxies are one of the key sources that have been considered as the origin of cosmic neutrinos, and catalogs were assembled for this purpose. Becker et al. \cite{2009arXiv0901.1775B} finds 127 starbursts within $z<0.03$, concluding that this sample should be close to complete. Futher, there are significant efforts under way to assemble galaxy catalogs in the local universe, for example, to aid gravitational wave observations. Some key catalogs in use include GWGC \cite{2011CQGra..28h5016W}, GLADE \footnote{\url{http://aquarius.elte.hu/glade/}} and CLU \cite{2016ApJ...820..136G}. Out to 100\,Mpc, these catalogs include galaxies with $\sim80$\% completeness \cite{2016ApJ...820..136G}. Since they mainly focus on star formation rate as a measure of the rate of compact binary mergers as key gravitational wave sources, star-forming and starburst galaxies are are likely cataloged at an even higher completeness within this range. Going out to 200\,Mpc, completeness in galaxies drops to $\sim40\%$. Nevertheless, the increased interest in gravitational wave follow-up observations guarantees significant efforts in assembling increasingly complete catalogs.

Current catalogs of AGN are currently less complete. For example, the latest SDSS quasar catalog contains only 232 quasars at $z<0.1$ within the surveyed $10^4$\,deg$^2$ \cite{2016arXiv160806483P}, corresponding to $\sim10\%$. A similarly limited number of AGN have been identified by the Swift-BAT 70-months survey, covering the entire sky \cite{2013ApJS..207...19B}.

Nearby blazars are expected to be bright and easily detectable. Available blazar catalogs include those detected by Fermi-LAT in high-energy $\gamma$-rays \cite{2016ApJS..222....5A}, with expected completeness $>80\%$ \cite{2013ApJS..206...12D}.

GCs are straightforward to identify at completeness even higher than galaxies due to their brightness. GC catalogs are nominally complete out to $z=2$ \cite{1989ApJS...70....1A}.

\subsection{Angular uncertainty}
\label{sec:angular}

One of the critical questions we aim to address here is how accurate neutrino angular reconstruction needs to be to ensure that a catalog-based search can identify the source population of cosmic neutrinos. We ran Monte Carlo simulations (Section \ref{section:MC}) to determine $\mathcal{P}(p_0)$ as a function of $\psi$ for different source types, using $T_{\rm obs}=10$\,yr with IceCube-Gen2, with $p_0 = 3\sigma$, and a threshold distance $d_{\rm th}=200$\,Mpc for all source types. Results are shown in Fig. \ref{figure:NeutrinoGalaxy}. We find that (i) blazar origin can be easily identified with a catalog-based search, as expected; for an angular precision of $\psi \sim 0.3^\circ$, (ii) AGN origin may be identified with $\mathcal{P}(p_0)=30-60\%$ probability, (iii) starburst galaxies could be identified with $\mathcal{P}(p_0)=50\%$, while (iv) GCs could be identified with $\mathcal{P}(p_0)=30\%$.

\begin{figure}
\resizebox{0.5\textwidth}{!}{\includegraphics{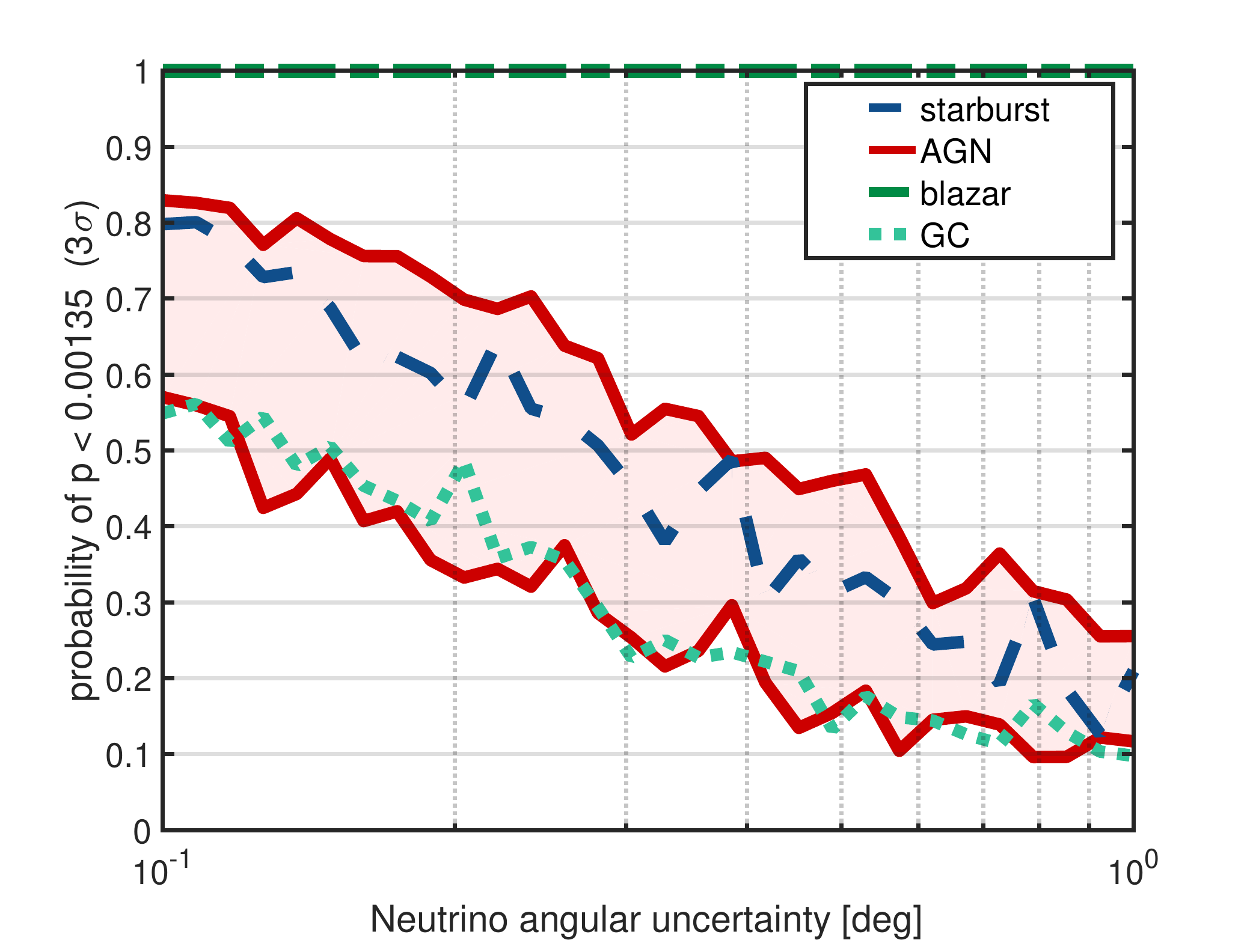}}
\caption{Probability of a signal+background observation having a p-value $\leq p_0=0.00135$ ($3\sigma$) as a function of neutrino angular uncertainty, for different source assumptions (see legend). For AGN, the spread corresponds to the range $10^{-5}-10^{-4}$\,Mpc$^{-3}$ source density. The results are obtained assuming $T_{\rm obs}=10$\,yr with IceCube-Gen2 in Sunflower 240 configuration, for the highest energies with $>24$\,TeV neutrinos, corresponding to a detection rate of 36 astrophysical and 36 atmospheric neutrinos per year.}
\label{figure:NeutrinoGalaxy}
\end{figure}

\subsection{Comparison to searches at TeV energies}

High-energy neutrino observatories can detect neutrinos over a uniquely wide energy range. In the case of IceCube-Gen2, including extensions such as PINGU, neutrinos will be detectable from $\sim 10$\,GeV to $\gg 1$\,PeV. This range can cover distinct emission processes, and feature varying sensitivity and background properties; therefore, neutrino searches typically focus on a different parts of the total available energy range.

Here, we examine \emph{low-threshold} ($\gtrsim1$\,TeV) neutrino searches that cover most of the sensitive energy band of the detector, with characteristic $\sim$\,TeV neutrino energies. We discuss the sensitivity of these searches in comparison to searches over the highest energy ($>24$\,TeV) neutrinos that we discussed previously (see Section \ref{section:detectionrate}).

Focusing on the highest energies has the advantage of high signal-to-noise ratio, given the very low background, and the somewhat better direction reconstruction than at lower energies. It yields, however, a relatively small overall number of signal neutrinos. Low-threshold searches benefit from having significantly larger number of signal neutrinos, but they need to overcome significant background contamination, given the soft atmospheric background spectrum.



\begin{figure}
\resizebox{0.5\textwidth}{!}{\includegraphics{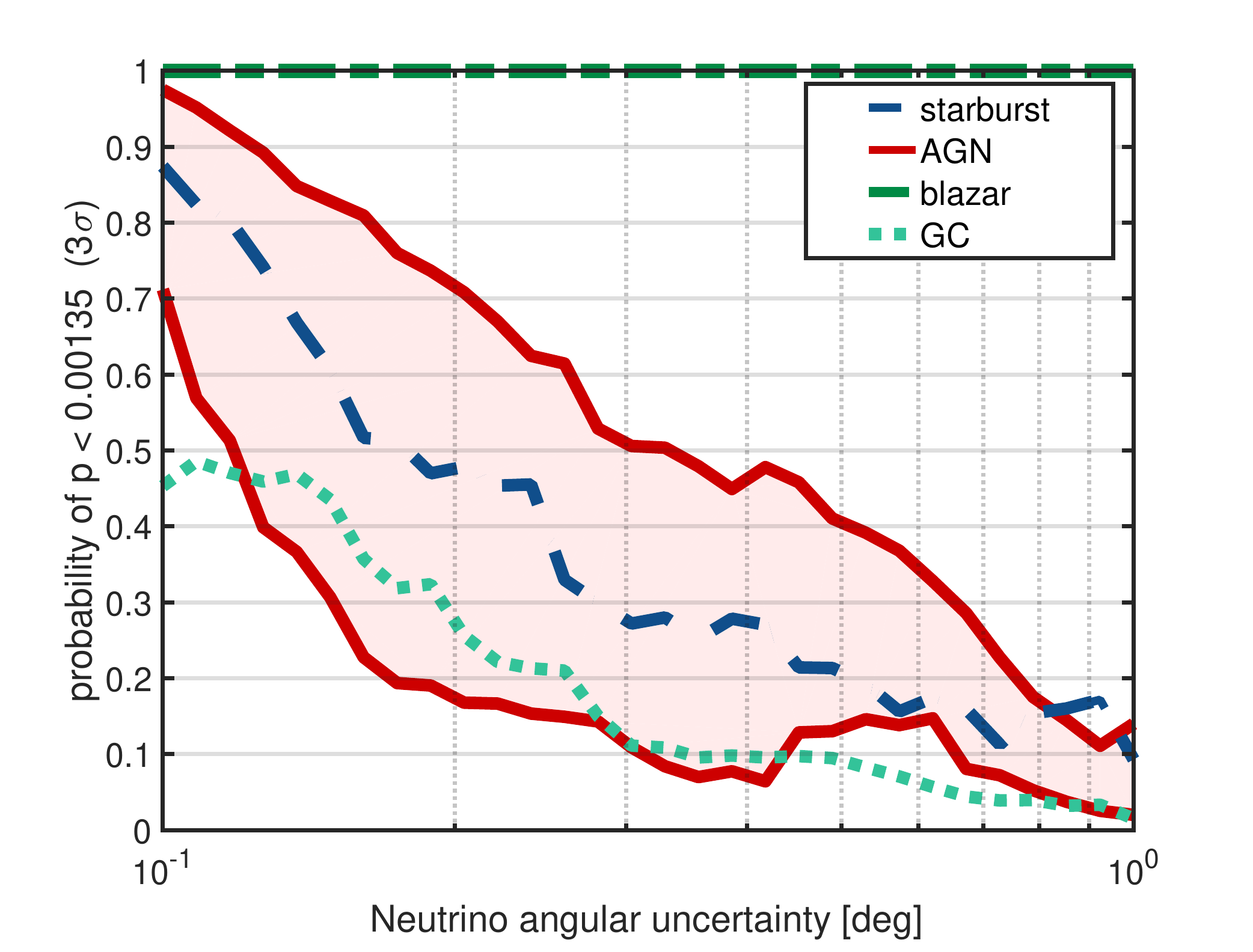}}
\caption{(Same as Fig. \ref{figure:NeutrinoGalaxy}, but for low-threshold neutrino search) Probability of a signal+background model having a p-value $\leq p_0=0.00135$ ($3\sigma$) as a function of neutrino angular uncertainty for a low-threshold ($>1$\,TeV) neutrino search, for different source models (see legend). The results are obtained assuming $T_{\rm obs}=10$\,yr with IceCube-Gen2.}
\label{figure:NeutrinoGalaxy_low-threshold}
\end{figure}

We repeat the Monte Carlo simulations used above for the case of the highest-energy neutrinos (Section \ref{sec:angular}), but now for the low-threshold case, to obtain the probability of identifying a source type as the origin of the astrophysical high-energy neutrinos. Results are shown in Fig. \ref{figure:NeutrinoGalaxy_low-threshold}. There are several points regarding the comparison of the low and high-threshold cases. (1) For larger angular uncertainty, the high-threshold sample performs better, while improving angular resolution results in the low-threshold case becoming more sensitive. This is intuitively expected. Once angular resolution is small enough with respect to the number of background overlaps (which depends on both the resolution and the source density), having more events helps and improves the sensitivity beyond the small sample available for the high-threshold case. (2) If neutrino energy probability distribution was used in the analysis--as they are in, e.g., standard IceCube point-source analyses \cite{2014ApJ...796..109A}, then a search could combine the benefits of both low and high-threshold analyses, since the contributions of the two cases would be distinguished, i.e. the result would always be at least as good as the result in either Figs. \ref{figure:NeutrinoGalaxy} and \ref{figure:NeutrinoGalaxy_low-threshold}.

We see that the probability of source type identification is comparable to that of the case of the highest-energy neutrinos, but is somewhat below it, especially for larger angular uncertainty. This result, of course, is strongly dependent on the assumed source spectrum. Softer source spectra, with fixed neutrino flux at the highest-energy range where astrophysical neutrinos have been detected, could strongly enhance the utility of $\sim$TeV neutrinos.

\subsection{Detector size}

\begin{figure}
\resizebox{0.5\textwidth}{!}{\includegraphics{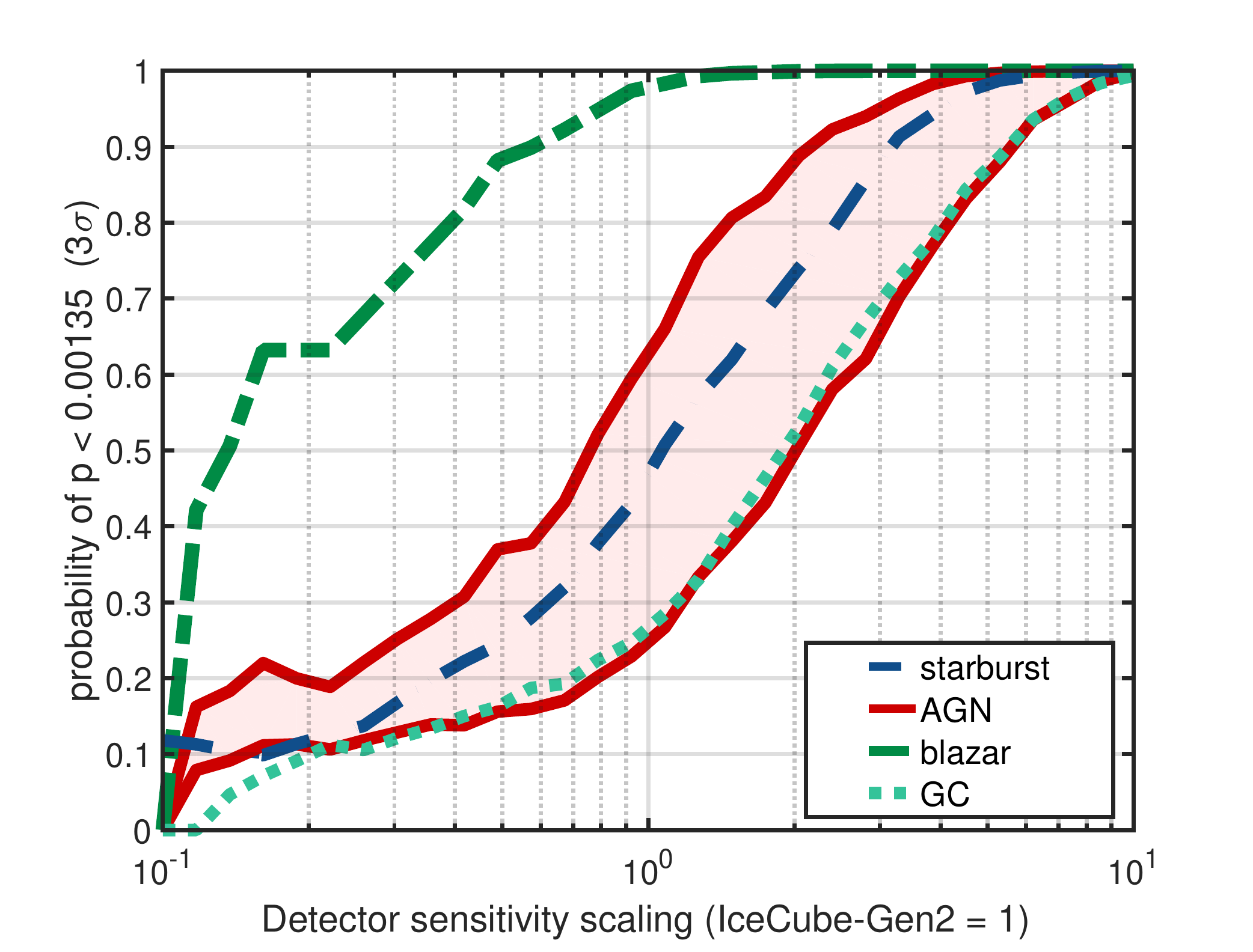}}
\caption{Probability of a signal+background observation having a p-value $\leq p_0=0.00135$ ($3\sigma$) as a function of the number of detected neutrinos. The x axis is normalized to the case of IceCube-Gen2 in Sunflower 240 configuration, for the highest energies with $>24$\,TeV neutrinos, which corresponds to a detection rate of 36 astrophysical and 36 atmospheric neutrinos per year. We show results for different source assumptions (see legend). The number of both astrophysical and atmospheric neutrinos is scaled identically. For AGN, the spread corresponds to the range $10^{-5}-10^{-4}$\,Mpc$^{-3}$ source density. The results are obtained assuming an angular uncertainty of $0.3^\circ$, and $T_{\rm obs}=10$\,yr.}
\label{figure:NeutrinoGalaxy_size}
\end{figure}

Beyond angular uncertainty, another variable of interest is the size of the detector, which largely determines sensitivity. To examine the role of detector size, we use Monte Carlo simulations to calculate the probability to identify a source type for different neutrino configurations. In particular, we characterize detector size by changing the detection rate of astrophysical and atmospheric neutrinos. We use an identical change for the two neutrino populations. The role of detection rate is shown in Fig. \ref{figure:NeutrinoGalaxy_size}, for a fixed angular uncertainty of $0.3^\circ$. We see that identically increasing the signal and background rate can significantly alter the chance of identifying the source type of origin. For starburst galaxies and AGN, identification probability around the IceCube-Gen2 in Sunflower 240 configuration is essentially proportional to the detection rate, signifying a rapid change.

\subsection{Veto power}

\begin{figure}
\resizebox{0.5\textwidth}{!}{\includegraphics{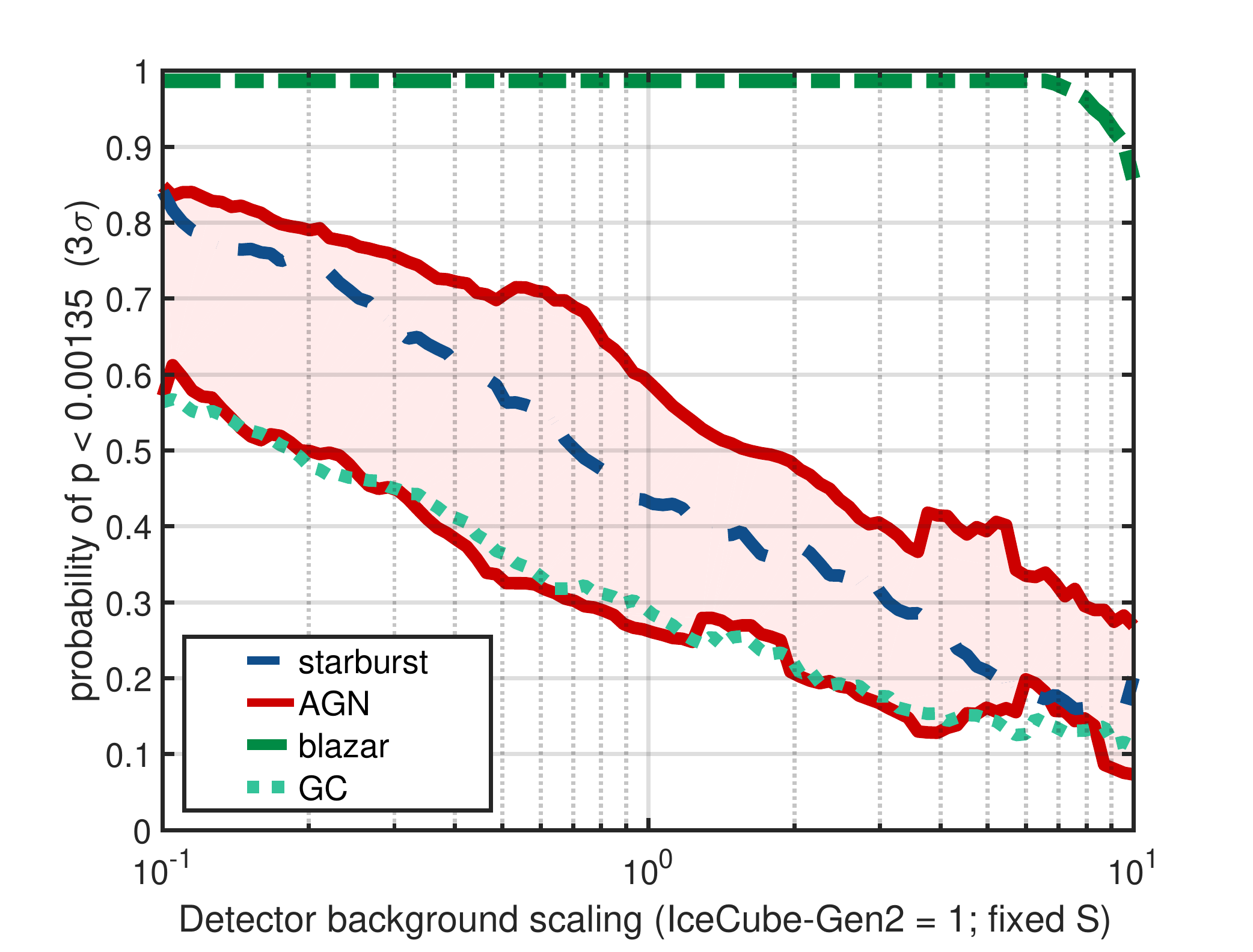}}
\caption{Probability of a signal+background observation having a p-value $\leq p_0=0.00135$ ($3\sigma$) as a function of the number of detected atmospheric neutrinos. The x axis is normalized to the case of IceCube-Gen2 in Sunflower 240 configuration, for the highest energies with $>24$\,TeV neutrinos, which corresponds to a detection rate of 36 atmospheric neutrinos per year. The detection rate of astrophysical neutrinos is kept fixed at 36\,yr$^{-1}$. We show results for different source assumptions (see legend). For AGN, the spread corresponds to the range $10^{-5}-10^{-4}$\,Mpc$^{-3}$ source density. The results are obtained assuming an angular uncertainty of $0.3^\circ$, and $T_{\rm obs}=10$\,yr.}
\label{figure:NeutrinoGalaxy_veto}
\end{figure}

While detector size may similarly affect the number of detected astrophysical and atmospheric neutrinos, it is also possible to improve capabilities to identify and filter out a fraction of atmospheric neutrinos, for example via surface arrays. Here, we examine the utility of such vetoes by considering the change in the probability to identify the neutrino's source of origin as a function of the atmospheric neutrino rate. We take the case of IceCube-Gen2 in Sunflower 240 configuration for the highest energies with $>24$\,TeV neutrinos, keep the detection rate of astrophysical neutrinos fixed, and vary the detection rate of atmospheric neutrinos. Results are shown in Fig. \ref{figure:NeutrinoGalaxy_veto}, for a fixed angular uncertainty of $0.3^\circ$. As expected, we find that the identification probability strongly depends on the atmospheric neutrino detection rate. For starburst galaxies and AGN, a factor of two decrease (increase) in the number of detected, un-vetoed atmospheric neutrinos increases (decreases) identification probability by $\sim20$\%, similarly to the utility of detector size.

\subsection{Locally suppressed starburst contribution}
\label{sec:suppression}

In Section \ref{section:fraction} we discussed the possibility of starburst galaxies having a steeper cosmic evolution than star formation rate, which would suppress the fraction of local neutrino emission. While the latest infrared luminosity function measurements by Herschel find that this suppression is not significant \cite{2013MNRAS.432...23G,2014JCAP...09..043T}, here we comment on how the results would be different if such a steeper evolution occurred.

For a factor of $\sim10$ reduction in the local contribution of starburst galaxies, the consequence is an effective 10-fold reduction in the number of astrophysical neutrinos from nearby sources, while the total number of detected neutrinos remains unchanged---neutrinos from more distant sources will be considered background for the purposes of source-catalog-based searches. Following the analytical estimate of Bartos \cite{2016APh....75...55B}, this should correspond to a factor of $10$ decrease in detection probability compared to the unsuppressed case. We carried out the Monte-Carlo analysis for the highest energy neutrinos as presented in Section \ref{section:MC}, with $f_{0}^{\rm suppressed} \approx 1.3\times10^{-5}$ instead of $f_{0}^{\rm sfr} \approx9\times10^{-5}$. We find that the factor of 10 decrease is a reasonable approximation. With an angular uncertainty of $0.3^\circ$, we find a detection probability for starburst galaxies with suppressed local contribution to be $\sim10\%$.

\section{Conclusion}
\label{section:conclusion}

We examined the prospects of utilizing the anisotropy of possible neutrino-sources to probe the origin of the cosmic high-energy neutrinos. We considered the use of catalogs of starburst galaxies, AGN and blazars, and explored the role of the angular uncertainty of the detector, as well as the distance reach of a source catalog, in identifying the source of the neutrinos.  Since we focused on continuous sources, we relied on neutrino track events with significantly better angular resolution than cascade events.

We presented a search method that can efficiently incorporate the directional and distance information of sources in a catalog-based search that does not rely on neutrino-neutrino coincidence. We were particularly interested in starburst galaxies and AGN since their number density is high enough that they are considered difficult to resolve with neutrinos \cite{2014PhRvD..90d3005A}. We looked at next generation neutrino detectors such as IceCube-Gen2 with significantly increased sensitivity and angular resolution compared to current detectors. Our main findings are as follows:
\begin{enumerate}
\item \textbf{Identification of challenging source types:} Our main conclusion is that it may be feasible to identify the source population of origin of high-energy neutrinos even for some of the challenging source types such as starburst galaxies or AGN using next generation high-energy neutrino observatories. Strong cosmic evolution, nevertheless, reduces the probability of source identification.
\item \textbf{Source catalogs are beneficial:} Search sensitivity using a source catalog exceeds that of using only the closest sources, and beyond some threshold distance, extending the source catalog further brings only marginal improvement.
\item \textbf{Required catalog depth}: We find that, for typical angular resolutions $\psi\gtrsim0.3^\circ$, it is viable to have a complete source catalog out to $\sim100$\,Mpc. A complete catalog out to this distance cannot be significantly improved by going to higher distances, except if high accuracy direction reconstruction is available ($\psi\lesssim0.1^\circ$), or for lower number-density source types such as blazars. This is good news, since it is difficult to assemble complete source catalogs for distances much farther than 100\,Mpc for most source types.
\item \textbf{Role of detector characteristics:} We examined the role of the detector's (i) angular resolution, (ii) size and (iii) veto capability in source identification. We characterized results in comparison to IceCube-Gen2 in Sunflower 240 configuration with a search for the highest energies with $>24$\,TeV neutrinos, with typical angular uncertainty of 0.3$^\circ$. Size was assumed to identically change $R_{\rm \nu,ast}$ and $R_{\rm \nu,atm}$, while veto capability changes $R_{\rm \nu,atm}$ only. We find that, for angular resolution and detector size, a factor of 2 change corresponds to a change in identification probability, $P$, of roughly 20\%. The dependence is somewhat weaker for veto power. These results are consistent with the analytical estimates of Bartos \cite{2016APh....75...55B}, who finds the following scaling relationship for the probability that the source of high-energy neutrinos will be identified:
    \begin{equation}
    P \propto \Psi^{-2/3}R_{\rm \nu,ast}(R_{\rm \nu,ast}+R_{\rm \nu,atm})^{-1/3}.
    \end{equation}
\end{enumerate}

The utilization of source catalogs can be further improved over the method presented above. For example, the expected neutrino luminosity for each source can be incorporated in the anisotropy. Optimally, one needs to weight each source with their anticipated neutrino flux. The angular distance of the neutrino and source directions can also be explored as a probability density functions. Further, the incorporation of neutrino energy probability densities could improve sensitivity and combine the benefits of the low and high-threshold searches presented here. Such changes will further improve the prospects of using source catalogs. Additionally, currently existing catalogs for some source types, primarily AGN, are highly incomplete, making the assembly of deep and complete source catalogs critical to maximize the reach of multimessenger studies. This may be the most feasible by focusing only on the directions of the detected neutrinos \cite{2041-8205-801-1-L1}.

\vspace{4 mm}
We thank Marek Kowalski and Kohta Murase for useful suggestions and Jakob van Santen for his help with IceCube-Gen2's expected detection rates. IB and SM are thankful for the generous support of Columbia University in the City of New York. MA and CF are grateful for the support of the Swedish Research Council, Stockholm University, and the Oskar Klein Centre.

\bibliographystyle{h-physrev}

\end{document}